\documentstyle[epsfig,twoside,12pt]{article}
{\catcode `\@=11 \global\let\AddToReset=\@addtoreset}
\AddToReset{equation}{section}

\def\greaterthansquiggle{\raise.3ex\hbox{$>$\kern-.75em\lower1ex\hbox{$\sim$}}}
\def\lessthansquiggle{\raise.3ex\hbox{$<$\kern-.75em\lower1ex\hbox{$\sim$}}}

\newcommand{\beq}{\begin{equation}}
\newcommand{\eeq}{\end{equation}}
\newcommand{\beqa}{\begin{eqnarray}}
\newcommand{\eeqa}{\end{eqnarray}}
\newcommand{\beqan}{\begin{eqnarray*}}
\newcommand{\eeqan}{\end{eqnarray*}}
\newcommand{\ba}{\begin{array}}
\newcommand{\ea}{\end{array}}
\newcommand{\no}{\nonumber}

\newcommand{\Tr}{\rm Tr\,}

\newcommand{\ol}{\overline}
\newcommand{\ra}{\rightarrow}

\newcommand{\vp}{\varphi}

\newcommand{\dg}{\dagger}

\newcommand{\A}{{\cal A}}

\newcommand{\Ha}{{\cal H}}
\newcommand{\dfrac}{\displaystyle \frac}

\newcommand{\dsum}{\displaystyle \sum}

\def\nz{\ifmmode {I\hskip -3pt N} \else {\hbox {$I\hskip -3pt N$}}\fi}
\def\zz{\ifmmode {Z\hskip -4.8pt Z} \else
       {\hbox {$Z\hskip -4.8pt Z$}}\fi}
\def\qz{\ifmmode {Q\hskip -5.0pt\vrule height6.0pt depth 0pt
       \hskip 6pt} \else {\hbox
       {$Q\hskip -5.0pt\vrule height6.0pt depth 0pt\hskip 6pt$}}\fi}
\def\rz{\ifmmode {I\hskip -3pt R} \else {\hbox {$I\hskip -3pt R$}}\fi}
\def\cz{\ifmmode {C\hskip -4.8pt\vrule height5.8pt\hskip 6.3pt} \else
       {\hbox {$C\hskip -4.8pt\vrule height5.8pt\hskip 6.3pt$}}\fi}

\normalsize
\begin{document}
\bibliographystyle{plain}
\begin{titlepage}
\begin{flushright}
UWThPh-2003-37\\
December 19, 2003
\end{flushright}

\vspace*{2.7cm}
\begin{center}
{\Large \bf
Finite supersymmetry transformations}\\[60pt]

{\sl Nevena Ilieva $^{a,c,*}$, \,Heide Narnhofer $^{b}$ and
Walter Thirring $^{b,c}$ }\\[30pt]

\noindent {\small $^a$ Theory Division, CERN}\\
\noindent {\small $^b$ Institut f\"ur Theoretische Physik,
Universit\"at Wien}\\

\noindent {\small $^c$ Erwin Schr\"odinger International
Institute for Mathematical Physics, Vienna}\\

\vspace{1.6cm}

\begin{abstract}

We investigate simple examples of supersymmetry algebras with real
and Grassmann parameters. Special attention is payed to the finite
supertransformations and their probability interpretation.
Furthermore we look for combinations of bosons and fermions which
are invariant under supertransformations. These combinations
correspond to states that are highly entangled.

\end{abstract}
\end{center}

\vspace{2.6cm}

{\footnotesize

$^\ast$ On leave from Institute for Nuclear Research and Nuclear
Energy, Bulgarian Academy of Sciences, Boul.Tzarigradsko Chaussee
72, 1784 Sofia, Bulgaria}


\vfill
\end{titlepage}
\setcounter{page}{2}

\section*{Introduction}

Supersymmetry is well understood and widely exploited in QFT and in
models as a set of infinitesimal transformations generating some
essential selection rules (see, e.g. \cite{WB}. The step towards
obtaining the finite
supertransformations is usually considered as unneccessary, thus placing
SUSY in a distinguished position as compared with the other symmetries we
know. On the other hand, already the simplest supersymmetric model -- SUSY
QM \cite{W1}, has a consistent physical interpretation in completely
conventional means as it is equivalent to one particle and one spin, a
system for which e.g. the probability interpretation perfectly holds. This
observation motivates one to inquire how far is it possible to construct a
realization of SUSY that avoids the introduction of Grassmann parameters
and thus allows for a probability interpretation. Some efforts in this
direction are due to Levine and Tomozawa \cite{LT} who tried doing this at
the price of additional generators in the (extended) Lie algebra.

We choose another strategy, namely we consider a graded $^*$-algebra $\A$
which is generated by some bosonic and fermionic operators and
possibly some Grassmann or Clifford variables. As supertransformations
we take $^*$-automorphisms of $\A$ which do not
respect the grading, thus mix bosons and fermions. The only other
structural element we use is a one-parameter group of automorphisms of
$\A$ (the time evolution) which commutes with the supertransformations.
The emphasis of this paper is different from the one most commonly seen in
the wast literature on this subject ($\sim 10^{4.5}$ papers) in the following
respects:
\begin{itemize}
\item[{\sl(i)\/}] We consider the supersymmetry as a symmetry in the
  conventional understanding. Thus we do not stay on the Lie-algebraic
  level but consider finite supertransformations. Therefore for our
  operators the product is
also defined and not only the commutator or the anticommutator;
\item[{\sl(ii)\/}] We do not require that the time evolution
is part of the Poincare-group which is represented by automorphism groups
of $\A$;
\item[{\sl(iii)\/}] We want to work with the standard probability
  interpretation of quantum mecha\-nics. There a state over a
  $^*$-algebra gives a probability distribution and a
  representation. The case II of Grassmann variables (defined in
  Sec.2) becomes then entirely strange. In the representation there
  are vectors of zero norm and all transition probabilities in this
  sector vanish.
\end{itemize}

In this context, two important questions to be answered are the
following: what happens under such supersymmetry transformations
with the states and which are the invariant structures. The former
is crutial for the probability interpretation of the theory, for
the latter, an example of an invariant state is certainly provided
by the Fock vacuum. However, the interplay between supersymmetry
and temperature and its consequences for the particle physics  are
less trivial and are still not matter of consense.

Some of these issues we could find only briefly discussed in the
literature (see, e.g. \cite{WB,PF}), others, up to our knowledge,
have not yet received the due attention. In particular, the
question whether the usual thermal state over one bosonic and one
fermionic mode breaks the supersymmetry with Grassmann variables
is confusing because of the seemingly contradictory results to be
found in the literature. Girardello et al. \cite{GGS} concluded
that SUSY is broken because only the ground state is annihilated
by the supercharges and the higher states are not. Though this
statement is correct, van Hove calculated carefully all
contributions to the change of the state to first order in the
superparameter and showed that they cancel \cite{vH}. Thus he
concluded that SUSY was not broken. Again this conclusion is
incorrect, as we find that for the supercharges there is a change
in the state (Sec. 6). Then came the sweeping proof of Buchholz
and Ojima \cite{BO} that the KMS-properties and SUSY are
incompatible even to first order. This proof seems impeccable
since it just uses the fact that a supersymmetric Hamiltonian is
positive but for a KMS-state its spectrum is $\bf R$. This appears
in the thermodynamic limit. Here we concentrate on finitely many
modes of fermions and bosons and therefore the spectrum is
semibounded and discrete. Already on this level we will see that
odd derivations violate the invariance of the thermal state.


Our result is that supertransformations can very well be constructed
without the help of Grassman variables. Moreover, when Grassmann
variables are involved, the supertransformations change the thermal
state, while in the case of real variables they leave it invariant.

\medskip

\section{$^*$-automorphisms and supertransformations}

The problems addressed in what follows are matters of principle and
appear already in the simplest supersymmetric situation of one Bose-
and one Fermi- mode. There we have the
supersymmetry as a one-parameter group of transformations of the algebra
$\A$ generated by the two creation and annihilation operators. In
fact, we shall consider three different mixed Bose--Fermi algebras --- $\A$,
$\A_\theta$ and $\A_C$, defined as follows: \footnote{To be conform with
the common notations in the supersymmetry literature we denote the hermitian
conjugation by $^\dg$ but we keep the $^*$-terminology for the algebraic
considerations.}
\begin{itemize}
\item[{\bf I.}] $\A$ is the algebra generated by the operators
$a, a^\dg, b, b^\dg$, satisfying CAR, resp. CCR
\beq \ba{ccccl}
\{a,a^\dg \} &=& [b, b^\dg] &=& 1 \\[4pt]
[\,a,b\,] &=& [a, b^\dg] &=& 0\,.
\ea \eeq
\item[{\bf II.}] The algebra $\A_\theta$ is the same $\A$, extended by a
Grassmann variable $\theta$ so that
\beq
\theta = \bar\theta, \quad \theta^2 = \{\theta, a\} =
[\theta, b] = 0. \eeq
\item[{\bf III.}] Furthermore we consider the case of Clifford
  variables  where we have
$\{\theta,\bar\theta\}=1$, the corresponding algebra being denoted by
$\A_C$. Its odd elements contain odd powers of the fermionic
operators $a, a^\dg$.
\end{itemize}

\paragraph{Remark}
$\A$ and $\A_C$ are $C^*$-algebras, whereas $\A_\theta$
is not, since the $C^*$-condition $\Vert
\bar\theta\theta \Vert = \Vert \theta\Vert^2$ would imply
$\Vert\theta\Vert=0$,
therefore $\theta=0$. $\A_\theta$ is
however a $^*$-algebra and a Grassmann module; any of its elements can be
written as $A+\theta B$ \cite{GP}. The ``soul'' $\theta B$ is a two-sided
ideal of $\A_\theta$, the ``body'' $\A$ being the quotient
algebra. The Hilbert-space representations are not faithful, in them
the soul vanishes. In such a representation $e^{isG_\theta}=1$ and the
supertransformations become an illusion.

\smallskip
A unitary element $U, \, U^\dg=U^{-1}$, creates a transformation of
the algebra which preserves both the multiplicative and the
$^*$-structure. We are interested in one-parameter groups of
supertransformations which preserve the total particle number
$N=a^\dg a + b^\dg b =: N_f + N_b$ but not the individual $N_f$
and $N_b$. The simplest generators for such transformations for
the algebras under consideration are the hermitean elements
\beq\ba{lcl}
G = ab^\dg + a^\dg b, & \qquad & \mbox{for }\, \A\\[4pt]
G_\theta = -i\theta(ab^\dg + a^\dg b) = G_{\bar\theta}\,,& & \mbox{for }
\,\A_\theta\\[4pt]
G_{\theta,\bar\theta} = 
-i(\theta a b^\dg + \bar\theta a^\dg b)\,,& &
\mbox{for }\,\A_C\ea
\eeq
Thus
\beq
A \ra A(s) = e^{iGs}A\,e^{-iGs},\quad A\in\A
\eeq
gives a $^*$-automorphism group of $\A$ and similarly for $\A_\theta$.
From the definitions follows $G^2=N$ and $G_\theta^2=0$. Thus
\beq e^{iGs} = \cos {\sqrt{N}s} + \frac{i G}{\sqrt N} \sin {\sqrt{N}s}
\eeq
and
\beq
e^{iG_\theta s} = 1 + \theta\,G\,s\,.
\eeq
\paragraph{Remarks}
\begin{enumerate}
\item
Alternatively we could have considered $G_\A:=i(a^\dg b - ab^\dg)$ but
this is equivalent to $G$ since the two are related by an even
transformation commuting with the time evolution $e^{iNt}\,:$  $G_\A =
e^{iN_f\pi/2}G e^{-iN_f\pi/2}$. However we shall later find a time
evolution which is still supersymmetric in the sense that it leaves
$G$ invariant but no longer $G_\A$;
\item As another alternative, one can consider the group generated by
  both $G$ and $G_\A$. This group is infinite-dimensional since $[G,\,
  G_\A] = 2i(b^\dg b - a^\dg a - 2a^\dg ab^\dg b)$ is not linearly
  expressible by these generators. In contradistinction, the
  corresponding generators with Grassmann parameters --- $G_\theta$
  and $G_{\A\theta}$, form a two-dimensional group;
\item The group generated by $G$ and $N$ is just the product of the
  two groups and thus isomorphic to ${\bf R^2}$, in sharp contrast
  to the Lie superalgebra. The only trace on the level of finite
  transformations of the relation between $N$ and $G$ is the simple
  expression (1.5) for $e^{iGs}$.
 \end{enumerate}

\medskip
Thus we conclude that there are four different candidates for possible SUSY
transformation of the three algebras:

\medskip
\noindent{\bf Ia.} In $\A$ one can generate a one-parameter group
of automorphisms by \beq A(s) = e^{iGs}Ae^{-iGs}, \quad G=a^\dg b
+ ab^\dg, \, A\in\A; \eeq In this spirit, for instance, the
Jaynes--Cummings model has been investigated in \cite{GL}.

\medskip
\noindent{\bf Ib.} In $\A$ an odd derivation generates a
one-parameter group of linear transformations by \beq A'=
\frac{d}{ds}A(s) := \delta A, \quad \delta A = \left\{\,
\ba{ll} \{G, A\} & \mbox{ for } A \mbox{ odd }\\[4pt]
i\,[G, A] & \mbox { for } A \mbox { even } \ea \right. \eeq where
$A$ is the basis $(a,b)$ of $\A$ and $\delta$ is such that $\delta
A^* = (\delta A)^*$. Thus one can express supertransformations
infinitesimally without Grassmann or Clifford variables (see, e.g.
\cite{BO,B2}). We take the convention with the factors $i$ such
that $\delta$ is compatible with conjugation, $\delta A^* =
(\delta A)^*$. However, $\delta$ is not compatible with
multiplication since only for even elements $b_i$ it satisfies the
Leibnitz rule $\delta(b_1 b_2) = (\delta b_1)b_2 +b_1 \delta b_2$.
For odd elements $a_i$ we get $\delta (ab) = (\delta a)b +
ia(\delta b)$ and $\delta(a_1 a_2) = i(\delta a_1)a_2 -ia_1\delta
a_2$. Therefore it can be integrated to a one-parameter group of
maps $\A \ra \A$ which commute with the time evolution and respect
the linear and $^*$-structure of $\A$ but not its multiplicative
structure. Note that for any element $A \in \A$, $\,\delta^2 A =
\delta\delta A = i\,[G^2,A]$, which gives its time derivative if
$G^2 = H$.

\medskip
\noindent{\bf II.} In $\A_\theta$ a one-parameter group of
automorphisms is generated by
$$
A(s) = e^{iG_\theta s}A e^{-iG_\theta s}, \quad G_\theta = -i\theta G
= G^\dg_\theta, \, A\in\A_\theta;
$$
\noindent{\bf III.} In $\A_C$ the operator $G_{\theta, \bar\theta} =
-i(\theta ab^\dg + \bar\theta a^\dg b) = G^\dg_{\theta, \bar\theta}$
in turn generates a one-parameter automorphism group
$$
A(s) = e^{i G_{\theta,\bar\theta}s}Ae^{-iG_{\theta,\bar\theta}s}, \,\,
A\in \A_C\,.
$$
Such a transformation becomes of importance in the noncommutative
supersymmetric models, which are defined by algebraic structure of this
type (see, e.g. \cite{S1}).

\medskip
Note that in the last two cases we are faced with automorphisms of
the extended algebras which do not leave $\A$ as a set invariant.

\medskip

\section{Finite supertransformation and differential characterization of the
groups}

{\bf Ia.\,} In order to obtain from Eq.(1.5) the finite
supertransformations in $\A$ explicitly one has to make use of the relations
$$
Na=a(N-1)\,, \quad Nb=b(N-1)
$$
$$
Ga=b-aG\,, \quad Gb=-a+bG
$$
to rearrange the products of
$$
a(s) = \left(\cos {\sqrt{N}s} +
\frac{i G}{\sqrt N} \sin {\sqrt{N}s}\right)a\left(\cos {\sqrt{N}s}
- \frac{i G}{\sqrt N} \sin {\sqrt{N}s}\right)
$$
and
$$
b(s) = \left(\cos {\sqrt{N}s} + \frac{i G}{\sqrt N} \sin
{\sqrt{N}s}\right)b\left(\cos {\sqrt{N}s} - \frac{i G}{\sqrt N}
\sin {\sqrt{N}s}\right)\,.
$$
This gives
\beqa
a(s) & = & \cos{\sqrt{N}s}\cos{\sqrt{N+1}s}\,a +
\frac{\sin{\sqrt{N}s}\sin{\sqrt{N+1}s}}{\sqrt{N(N+1)}}\,a^\dg
b^2\no\\[4pt]
& + & i\,\frac{\sin{\sqrt{N}s}\cos{\sqrt{N+1}s}}{\sqrt{N}}\,aa^\dg b -
i\,\frac{\sin{\sqrt{N+1}s}\cos{\sqrt{N}s}}{\sqrt{N+1}}\,a^\dg a b\\[4pt]
b(s) & = & \cos{\sqrt{N}s}\cos{\sqrt{N+1}s}\,b +
\frac{\sin{\sqrt{N}s}\sin{\sqrt{N+1}s}}{\sqrt{N(N+1)}}\,(b^2b^\dg -
2aa^\dg b)
\no\\[4pt]
& + & \,i\frac{\sin{\sqrt{N}s}\cos{\sqrt{N+1}s}}{\sqrt{N}}\,(ab^\dg b +
a^\dg b^2)\\[4pt]
& - & i\,\frac{\sin{\sqrt{N+1}s}\cos{\sqrt{N}s}}{\sqrt{N+1}}\,(abb^\dg +
a^\dg b^2)\no
\eeqa
with
$$
a \equiv a(0)\, \qquad b \equiv b(0)\,.
$$

\smallskip\noindent
{\bf Ib.\,} For the simple algebra at hand one gets $\,\delta
a = b$ and $\delta b = -ia$, which can be integrated to
\beq
\ba{ccl}
a(s) &=& a\cos{s\sqrt i} + (b/\sqrt i)\, \sin{s\sqrt i}\\[4pt]
b(s) &=& b\cos{s\sqrt i} - a\,\sqrt{i}\,\sin{s\sqrt i}
\ea
\eeq

\smallskip\noindent
{\bf II.\,} The supertransformation in $\A_\theta$ is
particularly simple:
\beq
\ba{ccccccc} a(s) &=& a + s\theta b\,, &\quad& a^\dg(s) &=& a^\dg -
s\theta
b^\dg\\[6pt] b(s) &=& b - \theta s a\,, &\quad& b^\dg(s) &=& b^\dg +
\theta s a^\dg\ea
\eeq
$$
\theta(s) = \theta = \bar\theta(s)\,.
$$
One readily verifies that this is a $^*$-automorphism.

\medskip\noindent
{\bf III.\,} For $G_{\theta,\bar\theta}^2$ there is no simple
expression and thus no
way to get explicite expressions for $a(s)$ and $\,b(s)$.

\medskip
Alternatively we could
first look at the differential equation which
determines the flow. With the notation $A'= d/ds\,A(s)$ we have for the
formal derivatives:

$$\ba{rclc}
{\bf Ia.} & & \left\{\,
\ba{l}
a'=-ib + 2iGa\\[4pt]
b'=-ia
\ea \right. & \\[14pt]
{\bf Ib.} & \qquad &
\left\{\,\ba{lcl}
a' &=& b  \\[4pt]
b' &=& -ia
\ea \right. &\quad\\[14pt]
{\bf II.} & &
\left\{\,\ba{lcl}
a' = \theta b\,, &\quad& \theta' = 0\\[4pt]
b' = -\theta a &&
\ea \right. & \\[14pt]
{\bf III.} & &
\left\{\,\ba{lcl}
a' = \bar\theta b\,, &\quad& \theta' = -ba^\dg\\[4pt]
b' = -\theta a\,, & & \bar\theta' = -ab^\dg
\ea \right. & \\
\ea
$$

\paragraph{{\large\bf Remarks:}}
\begin{enumerate}
\item Compatibility with the product structure and therefore with canonical
commutation or anticommutation relations (CCR, resp. CAR)
requires:
\begin{itemize}
\item[$(\alpha)\,$] $a^\dg{}' a + a^\dg a' + a'a^\dg + aa^\dg{}' = 0$
\item[$(\beta)\,$] $b'b^\dg + bb'^\dg - b'^\dg b - b^\dg b' = 0$
\item[$(\gamma)\,$] $a'b + ab' - ba' - b'a = 0$
\end{itemize}

\noindent
Whereas $(\beta)$ and $(\gamma)$ are always satisfied, $(\alpha)$ holds only
in cases {\bf Ia}, {\bf II}, {\bf III}.
\item In cases {\bf II}, {\bf III} the commutation relations require
  in addition:
\begin{itemize}
\item[$(\alpha)\,$] $\theta' a + \theta a' + a'\theta + a\theta' = 0$
\item[$(\beta)\,$] $\theta'\bar\theta + \theta\bar\theta' +
  \bar\theta'\theta + \bar\theta\theta' = 0$\,.
\end{itemize}
These conditions are satisfied only in case {\bf II}. Thus to say
that the $\theta$'s are just anticommutative numbers and are not
  changed by the transformation leads to inconsistencies.
\end{enumerate}

To integrate the differential equations poses different problems in
the four cases, although this way we could obtain again the finite
transformations, Eqs.(2.1)--(2.4).

\medskip

\section{Representations}

A representation $\pi$ is an isomorphism of the algebra with an
operator algebra in a Hilbert space, a $^*$-representation is a
$^*$-isomorphism, i.e. $\pi(AB)=\pi(A)\pi(B), \pi(A^*)=\pi(A)^*$.
Thus for $\A_\theta$ we can only have representations and not
$^*$-representations since operator algebras are $C^*$-algebras. The
GNS-construction leads at first to an inner-product space which
contains zero-norm vectors created by the soul. Passing to the Hilbert
space by quotioning them out we are left only with the body, the
soul-ideal being represented by zero. But for a physical
interpretation this procedure is unavoidable since the results of
measurements are real and not Grassmann numbers.
This seemingly purely mathematical distinction will have the
consequence that in a representation the transformation in
$\A_\theta$ has no probability interpretation. It will not give
transition probabilities.

An obvious representation of $\A$ is given by a quantum particle
with coordinates $(x,p)$ and one spin \cite{W1}:
\beq
a = \frac{\sigma_x-i\sigma_y}{2}\,,\qquad\qquad
b = \frac{x+ip}{\sqrt 2}\,.
\eeq
For $\A_\theta$ we have to use a second spin, described
by Pauli-matrices $\tau_k$ and set $\theta = \sigma_3\,(\tau_x\!-\!i\tau_y)/2$.
The usual Fock representation $\pi_F$ based on a ``vacuum" $|0\rangle$
with $a |0\rangle = b|0\rangle = 0$ appears to be the most
convenient framework for our considerations. In $\pi_F$ an
orthogonal basis is given by
$$
|n_f, n_b, n_g\rangle =
(a^\dg)^{n_f}\frac{(b^\dg)^{n_b}}{\sqrt{n_b!}}\theta^{\,n_g}|0\rangle\,,
\quad n_f, n_g = 0,1\,, \,\, n_b = 0,1,2,...
$$
For $\A$ we have $n_g = 0$, for $\A_\theta$ we have $\theta |0\rangle \not=
0$ but $\theta^2 |0\rangle = 0$. The action of $G$ is now rather
simple
\beq
\ba{lcl} G\,|0, n_b\rangle &=& \sqrt{n_b}\,|1, n_b-1\rangle\\[6pt]
G\,|1, n_b\rangle &=& \sqrt{n_b+1}\,|0, n_b+1\rangle\,,\ea
\eeq
thus
\beqan
e^{iGs}|0, n_b\rangle &=& \cos \sqrt n_b\, s|0, n_b\rangle +
i\sin\sqrt n_b\, s |1, n_b-1\rangle\\[6pt]
e^{iGs}|1, n_b\rangle &=& \cos \sqrt{n_b+1}\, s|1, n_b\rangle +
i\sin\sqrt{n_b+1}\, s |0, n_b+1\rangle\,.
\eeqan
The action of $e^{iG_\theta s}$ is even simpler
\beq\ba{rcl}
e^{iG_\theta s}|0, n_b, 0\rangle &=& |0, n_b, 0\rangle -
s\sqrt n_b |1, n_b-1, 1\rangle\\[4pt]
e^{iG_\theta s}|1, n_b, 0\rangle &=& |1, n_b,0\rangle -s\sqrt{n_b+1}\,|0,
n_b+1, 1\rangle\\[4pt]
e^{iG_\theta s}|n_f, n_b, 1\rangle &=& |n_f, n_b, 1\rangle\,.
\ea\eeq
Once again, for $\A_\theta$ $\pi_F$ is only a representation but cannot be
a $^*$-representation. As a consequence only $e^{iGs}$ but not
$e^{iG_\theta s}$ has a physical interpretation. $e^{iGs}$ changes
a boson into a fermion or vice versa. It does this with a
probability $\sin^2 \sqrt n_b\, s$, resp.  $ \sin^2 \sqrt{n_b+1}\,s$,
whereas it leaves the state unchanged with $\cos^2$ probability. On
the contrary, $e^{iG_\theta s}$ does nothing for $n_g=1$ and for
$n_g=0$ it leaves the state unchanged with weight $1$ and changes
$n_g$ and bosons into fermions (or vice versa) with a weight $n_b\,
s^2$, resp.  $(n_b+1)\,s^2$, times $\Vert
\,\theta\vert\,\rangle\,\Vert^2$ which is zero. Clearly, these weights
should not be interpreted as probabilities and we are forced to
conclude that the supertransformation in $\A_\theta$ is only an illusion, in
contradistinction to the supertransformation in $\A$.

If we give up the hermiticity of $\theta$, that is $\theta$ is no longer
hermitian but instead $\{\theta, \bar\theta\} = 1$, still keeping the
anticommutativity, the algebra $\A_\theta = \{a, b, \theta\}$ becomes the
$C^*$-algebra $\A_C$ and we have a Fock $^*$-representation $\pi_F$. It is
based on the vacuum $|0\rangle$, which is annihilated by $\{a, b,
\theta\}$. An orthogonal basis is given by
\beq
|n_f, n_g, n_b\rangle =
(a^\dg)^{n_f}\,(\bar\theta)^{n_g}\,\frac{(b^\dg)^{n_b}}{\sqrt{n_b!}}
|0\rangle\,, \quad n_f, n_g = 0,1; \,\, n_b=0,1,2,...
\eeq
We restrict ourselves to inspect the unitary implementer $e^{iGs}$ of the
supertransformations. The previous $G$ generalises to $G_{\theta,\bar\theta}
= \theta ab^\dg +
ba^\dg\bar\theta$ and $G_{\theta,\bar\theta}^2$ is a bit more complicated,
$G_{\theta,\bar\theta}^2 =
N_b(1-N_f)(1-N_g) + N_fN_g(1+N_b)$, $N_g = \bar\theta\theta$, $G^2 =
N/2$
for $N_g = 1/2$. Still it is diagonal in $\pi_F$ and there is nothing
wrong with the expansion (1.5), with $N$ replaced with $G^2$. To work it
out we need the action of $G_{\theta,\bar\theta}$:
\beq
\ba{lcl} G_{\theta,\bar\theta}|0, 0, n_b\,\rangle &=&
\sqrt n_b |1,1,n_b-1\rangle\\[7pt]
G_{\theta,\bar\theta}|1, 1, n_b\,\rangle &=&
\sqrt{n_b+1}|0, 0, n_b+1\rangle \ea
\eeq
Therefore
\beq
\ba{lcl} e^{iG_{\theta,\bar\theta}s}|0, 0, n_b\rangle &=&
\cos \sqrt n_b\,s|0, 0, n_b\rangle +
i \sin \sqrt n_b\, s|1, 1, n_b-1\rangle\\[7pt]
e^{iG_{\theta,\bar\theta}s}|1, 1, n_b\rangle &=&
\cos \sqrt{n_b+1}\, s |1, 1, n_b\rangle +
i\sin\sqrt{n_b+1}\,s|0, 0, n_b+1\rangle\,.
\ea
\eeq

Here we are dealing again with a $C^*$-algebra and a $^*$-
representation, so the transition probabilities add up to unity.
In fact they are identical to the ones we found in $\A$, $\theta$
acts like a fermion (``spurion'') attached to the original one and
does nothing exceptional. Therefore we should honestly declare that we
have two fermions and restore the symmetry between fermions and
bosons. As a side remark we shall show that going to a finite number
of bosonic and fermionic modes changes in the Fock representation very
little.

\medskip

\section{Some generalizations}

\subsection{The $N$-fermion/$N$-boson system}

The $N$-fermion/$N$-boson system is defined through the algebra $\A =
\{a_i, b_j\},\, i,j,k,... = 1,2, \dots, N$, with the usual rules
\beq
\ba{l} \{a_i, a_k^\dg\} = \delta_{ik} = [b_i, b^\dg_k]\\[6pt]
\{a_i, a_k\} = [b_i, b_k] = [a_i, b_k] = [a_i, b^\dg_k] = 0\,.
\ea
\eeq
We go straight to the question of the supertransformation,
\beq G = \sum_{i=1}^n k_i (a_ib^\dg_i + a^\dg_i b_i)\,.
\eeq
In
$G^2$ the quartic terms again reduce to quadratic expression
\beq
G^2 = \sum_{i=1}^n k_i^2 (a^\dg_i a_i + b^\dg_i b_i) =: H\,.
\eeq
We call it $H$ since it looks like a popular Hamiltonian. In $H$
the terms with different $i$ commute, in $G$ they do not. As a
consequence,
$e^{iGs} \not= 
\otimes_{k=1}^N e^{iG_k s}$,
but in the Fock
representation it is still managable. At the risk of boring the
experts we give bellow the relevant expressions for the
two-boson/two-fermion system ($N=2$) explicitly. In this case, in the
orthogonal basis
\beq
|n_{f_i}, n_{b_j}\rangle =
(a_1^\dg)^{n_{f_1}}\,(a_2^\dg)^{n_{f_2}}
\,\frac{(b_1^\dg)^{n_{b_1}}(b_2^\dg)^{n_{b_2}}}{\sqrt{n_{b_1}!\,n_{b_2}!}}
\,|0\rangle\,,\quad i,j=1,2
\eeq
$G$ acts as
\beqa 
G\,|0, 0, n_{b_1}, n_{b_2}\rangle &=& k_1 \sqrt{n_{b_1}}\,|1, 0,
n_{b_1}-1, n_{b_2}\rangle + k_2\sqrt{n_{b_2}}\, |0, 1, n_{b_1},
n_{b_2}-1 \rangle\no\\[6pt]
G\,|1, 0, n_{b_1}, n_{b_2}\rangle &=& k_1 \sqrt{n_{b_1}+1}\,|0, 0,
n_{b_1}+1, n_{b_2}\rangle + k_2\sqrt{n_{b_2}}\, |1, 1, n_{b_1},
n_{b_2}-1 \rangle\no\\[6pt]
G\,|0, 1, n_{b_1}, n_{b_2}\rangle &=& k_1 \sqrt{n_{b_1}}\,|1, 1,
n_{b_1}, n_{b_2}\rangle  + k_2\sqrt{n_{b_2}+1}\, |0, 0, n_{b_1},
n_{b_2}+1 \rangle\\[6pt]
G\,|1, 1, n_{b_1}, n_{b_2}\rangle &=& k_1 \sqrt{n_{b_1}+1}\,|0, 1,
n_{b_1}+1, n_{b_2}\rangle +  k_2\sqrt{n_{b_2}+1}\, |1, 0, n_{b_1},
n_{b_2}+1 \rangle\no 
\eeqa
from which we calculate the unitary action of $\,e^{iGs} = \cos{\sqrt H}s
+ i\,\frac{G}{\sqrt H}\sin{\sqrt H}s\,\,$ to be
\beqan
& e^{iGs}\,|0, 0, n_{b_1}, n_{b_2}\rangle = \cos{s\sqrt{k_1^2 n_{b_1} +
k_2^2 n_{b_2}}}\,|0, 0, n_{b_1}, n_{b_2}\rangle + \\[6pt]
&i\,\dfrac{\sin{s\,\sqrt{k_1^2 n_{b_1} + k_2^2 n_{b_2}}}}{\sqrt{k_1^2
n_{b_1} + k_2^2 n_{b_2}}}\,\left(k_1\sqrt{n_{b_1}}\,|1, 0, n_{b_1}-1,
n_{b_2}\rangle + k_2\sqrt{n_{b_2}}\,|0, 1, n_{b_1}, n_{b_2}-1 \right)\\[8pt]
& e^{iGs}\,|1, 0, n_{b_1}, n_{b_2}\rangle = \cos{s\sqrt{k_1^2(n_{b_1}+1)
+ k_2^2 n_{b_2}}}\,|1, 0, n_{b_1}, n_{b_2}\rangle + \\[6pt]
&i\,\dfrac{\sin{s\,\sqrt{k_1^2(n_{b_1}+1) + k_2^2 n_{b_2}}}}{\sqrt{k_1^2
(n_{b_1}+1) + k_2^2 n_{b_2}}}\,\left(k_1\sqrt{n_{b_1}+1}\,|0, 0,
n_{b_1}+1,n_{b_2}\rangle + k_2\sqrt{n_{b_2}}\,|1, 1, n_{b_1}, n_{b_2}-1
\right)\\[7pt]
&\dots
\eeqan

Note that the transition probabilities for the three outcomes add up to
unity.

In the general case of $N$ modes the orthogonal basis is given by
$$
|\{n_j\}, \{m_j\}\rangle = \prod_{i=1}^N
(a_i^\dg)^{n_i}(b_i^\dg)^{m_i}|0\rangle\,,\qquad
H|\{n_j\},\{m_j\}\rangle = E|\{n_j\},\{m_j\}\rangle
$$
with
$$
\Vert\,|\{n_j\},\{m_j\}\rangle\Vert = 1 \qquad \mbox{if }\, n_j =
0,1\,;\, m_k = 0,1,2,...
$$
The action of $G$ and of the unitary transformation it implements,
correspondingly become
$$
\ba{ccl}
G\,|\{n_j\},\{m_j\}\rangle &=& \dsum_{i=1}^N k_i (\sqrt{m_i+1}\,
|n_1,\dots,n_i-1,\dots, n_N, m_1,\dots,m_i+1,\dots,m_N\rangle \\[6pt]
&+& \sqrt m_i\,|n_1,\dots,n_i+1,\dots, n_N,
m_1,\dots,m_i-1,\dots,m_N\rangle)
\ea
$$
$$
\ba{ccl}
e^{iGs}|\{n_j\},\{m_j\}\rangle &=& \cos{\sqrt E\,s}
\,|\{n_j\},\{m_j\}\rangle + i\,\sin{\sqrt E\,s} \dsum_{i=1}^N
k_i/\sqrt E \\[6pt]
&\times& \left(\sqrt{m_i+1}\,|n_1,\dots,n_i-1,\dots, n_N,
m_1,\dots,m_i+1,\dots,m_N\rangle \right. \\[7pt]
&+& \left.\sqrt m_i\,|n_1,\dots,n_i+1,\dots,
n_N, m_1,\dots,m_i-1,\dots,m_N\rangle\right)
\ea
$$
Note that
\beq
\Vert\,|\,\,\rangle\,\Vert^2 = \cos^2\sqrt E\,s + \sin^2\sqrt
E\,s\,\sum_{i=1}^N k_i^2 \times \,\left\{
\ba{ccl}(m_i+1) &\,\,& \mbox{if }\quad n_i=1\\[3pt]
m_i & & \mbox{if }\quad n_i=0
\ea \right\}\,= 1\,.
\eeq

\medskip

\subsection{The poorman's Wess--Zumino model}

As a next generalization and a step towards the field-theory setting
let us consider a model of one Bose- and one Fermi- mode, but with an
interaction introduced through the following modification of the
supercharge:
$$
G = \tilde Q + \tilde Q^\dg 
$$
\beq
\ba{ccc}
Q^\dg\, = \,a^\dg b &\,\ra\,& \tilde Q^\dg = a^\dg(b+gb^\dg b)\\[4pt]
Q \, = \, b^\dg a &\,\ra\,& \tilde Q = (b^\dg + gb^\dg b)a
\ea
\eeq
with $\,g\,$ real. This is nothing else but a prototype of
the Wess--Zumino model \cite{WZ} and we have been dealing so far
with its free-theory limit ($\,g=0\,$). Such operators on loop
space have been considered in \cite{AJ}.

Again, the Hamiltonian is given by
\beq
H_g = G^2 = \{Q,\, Q^\dg\}\,,
\eeq
that is
\beq
H_g = H_0 + gH_0(b + b^+) - gb^\dg + g^2(b^\dg b)^2
\eeq
with $H_0 = N$. Expansions (1.5), (1.6) still hold (because of
Eq.(4.8)), as well as the conservation of
the supercharge $G$, $[G, H] = 0$.

There are many possibilities for the supercharges and with (4.8) we
can always generate a time evolution commuting with the
supertransformations.
However, already in this simple model it turns out that these
supertransformations with different charges generate an
infinite-dimensional algebra contrary to
the free case \cite{N} which appears to be in this context a lucky
exception.

\medskip

\section{Supersymmetric quasiparticles}

The transformation (2.1), (2.2) mixes $a$ and $b$ in a rather complicated
manner and the question arises whether a special combination $A$ is  left
intact so that $e^{iGs}Ae^{-iGs}$ produces only a phase factor
$e^{i\gamma s}A$. This means that the commutator with $G$ should
reproduce $A$. For a
polynomial in $a$ and $b$ this does not happen, commuting with $G$ keeps
increasing the degree of the polynomial in $b$. However, for a
nonpolynomial function $f(b^\dg b)$ this is not necessarily so and we shall
show now that even a simple choice allows to get for the transformed $A'$
a phase factor with $\gamma = \pm 1$. We just take $f$ real, continuous
and $f(x)=0 \quad \forall x\geq 1$. Then $bf(b^\dg b)= f(b^\dg b + 1)b$
since
$b(b^\dg b)^n = (b^\dg b+1)^n b \quad \forall n \in N$. But since $b^\dg
b \geq 0$, $f(b^\dg b +1)=0$ and $bf(b^\dg b) = f(b^\dg b)b^\dg =0$.
Denoting $aa^\dg f(b^\dg b)$ by $P_0$, $aP_0 = P_0 a^\dg = 0$, we claim
that
$$
[A_\pm, G]= \mp A_\pm, \quad \mbox{ where }
A_\pm := P_0(a\mp b)\,.
$$
In the Fock representation $P_0$ is the projection onto the vacuum.

\noindent{\bf Proof:}
$$
\ba{l}
G A_\pm = (ab^\dg + a^\dg b) P_0(a\mp b) = 0\\[4pt]
b b^\dg = 1 + b^\dg b \quad \Rightarrow \quad P_0bb^\dg = P_0 aa^\dg
=P_0\\[4pt]
A_\pm G = P_0(a\mp b)(ab^\dg + a^\dg b) = P_0 (\mp\,bb^\dg a + b) =
\mp A_\pm\,.
\ea
$$

\noindent{\bf Conclusion:}
$$
A_\pm(s) = e^{isG}A_\pm e^{-isG} = e^{\pm is}A_\pm ,\,\, \mbox {
thus } A^\dg_\pm A_\pm \mbox { is \, invariant}
$$

\noindent{\bf Question:} \,Is the bastard created by $A_{\pm}$ (the
``susino") a boson or a fermion?

\noindent{\bf Answer:} \,Though $(A_\pm(s))^2=0$, neither $[A_\pm,
A^\dg_\pm]$ nor $\{A_\pm, A^\dg_\pm\}$ equals 1. $A_\pm$ correspond to
elementary SUSY-excitations and by combining them we can construct the
invariants $A_\pm A^\dg_\pm$ and SUSY excitons $A_\pm A^\dg_\mp$ which
have a phase factor $\gamma = \pm 2$.

\paragraph{{\bf Remarks:}}
\begin{enumerate}
\item Under the time evolution with $H=G^2$ the susinos $A_\pm$
evolve like the bosons or the fermions, $A_\pm(t) = e^{it}A_\pm$, but the
situation could be as in the $K^0 - \ol{K^0}$ system: by a small
perturbation neither the boson nor the fermion are time-invariant but only
the susinos. Consider $H_\alpha = H + \alpha G + \alpha^2/4 = (G +
\alpha/2)^2$. Under its time evolution neither $a$ nor $b$ but only
$A_\pm$ changes just by a phase factor. Thus in a perfectly supersymmetric
situation physics may become quite unusual;
\item Under the supertransformation generated by $G_\A$, $A_\pm$
  ocsillate rigidly
$$
[A_\pm, G_\A] = \pm\,i A_\mp\,.
$$
More explicitly, with
$A_\pm (s) = e^{isG_\A}\,A_\pm\,e^{-isG_\A}$, $A' = d/ds\,A(s)$,
we get $A'_+ = A_-$, $A'_- = -A_+$ and therefore the oscillations
$$
\ba{ccl}
A_+(r) & = & \,\,\,\, A_+(0)\,\cos r + A_-(0)\, \sin r\\[4pt]
A_-(r) & = & -A_+(0)\,\sin r + A_-(0)\, \cos r\,;
\ea
$$
\item $A_{\pm}$ can be generalized to
\beq A_{(m,n)\,\pm} = (a^\dg {b^\dg}^{m-1} + {b^\dg}^m) P_0(a^n
b^{n-1} + b^n) \eeq with the properties
$$
(A_{(m,n)\,\pm})^\dg = A_{(n,m)\,\pm}, \quad
A_{(m,n)\,\pm}A_{(n,r)\,\pm}
= n \,A_{(m,r)\,\pm}
$$
$$
e^{isG}A_{(m,n)\,\pm}e^{-isG} = e^{\pm is(\sqrt n - \sqrt m)}A_{(m,n)\,\pm}\,.
$$
It can be interpreted as absorbing $n$ and creating $m$ susinos.
\item For the evolution (2.3) even a linear combination of $a$ and $b$
changes only by a common factor, there we have
$$
a(s)/\sqrt i + b(s) = e^{t\sqrt i}(a/\sqrt i + b);
$$
\item The other two supertransformations mix in some $\theta$'s and do
  not leave a function of $a$ and $b$ only invariant.
\end{enumerate}

The unitaries $e^{isG}$ have eigenvalues $e^{\pm is\sqrt n}$, for
$n=0$ the eigenvector is the vacuum $|0\rangle$ and for $n=1$ they
are $A^\dg_\pm |0\rangle$. For arbitrary $n$ their properties are
described by the following lemma which relates two main streams of
contemporary physics:

\begin{quote}{\bf Lemma:}\,
{\sl Except for $n=0$, in all other eigenvectors of the super
transformation $e^{isG}$ the bosons and the fermions of the
one-boson/one-fermion system are maximally entangled.}
\end{quote}

The Hilbert space of our system is the tensor product of the
fermionic and bosonic Hilbert spaces, $\Ha = \Ha_F \otimes \Ha_B$
and vectors which are not of the product form are called
entangled, i.e. the correlations they carry are of quantum and not
of classical origin. A convenient measure of the entanglement of a
vector $|\,\rangle$ in $\Ha$ is the entropy of the fermionic
density matrix $\rho_F = -\Tr _{\Ha_B}\, |\,\rangle\langle\,|\,$,
$\,\Tr _{\Ha_B}$ being the partial trace in $\Ha_B$, namely
$$
E = - \Tr _{\Ha_F} \rho_F\ln \rho_F \leq \ln 2\,.
$$
However $\rho_B$, the state reduced to the bosons, has the same
entropy as $\rho_F$. Fermions are thereby not prefered to bosons.

\smallskip
\noindent{\bf Proof:}

\noindent One verifies (compare {\bf Remark 3} above)
$$
(a^\dg b + b^\dg a)(|1, n-1\rangle \pm |0, n\rangle) = \pm\sqrt n
(|1, n-1\rangle \pm |0, n\rangle)\,,
$$
thus
$$
e^{isG}(|1, n-1\rangle \pm |0, n\rangle) = e^{\pm is\sqrt n} (|1,
n-1\rangle \pm |0, n\rangle)\,.
$$
Calculating the fermionic density matrix $\rho_F$ we find that in
all cases it corresponds to the tracial state, $\rho_F = 1/2$ and
thus $E = \ln 2$ for any $n$. Since the transformation with
$e^{isG_\A}$ is unitarily equivalent to the one with $e^{isG}$ by
a unitary that belongs to the Fermi subalgebra this does not
change the entanglement and the above statement holds also for its
eigenvectors.

The generalization to two modes is straightforward but the situation there
is somewhat different. With the notations of Section 4.1, $k_1=1$, $k_2=k$,
the supercharge and the Hamiltonian become
$$
\ba{l}
G = a_1^\dg b_1 + a_1b_1^\dg + k\,(a_2^\dg b_2 + a_2 b_2^\dg)\\[4pt]
H = G^2 = a_1^\dg a_1 + b_1^\dg b_1 + k^2\,(a_2^\dg a_2 + b_2^\dg b_2)
\ea
$$
The eigenvalues of $H$ are $n_{f_1}\!+\!
n_{b_1}\!+\!k^2(n_{f_2}\!+ \!n_{b_2})$, correspondingly those of
$G$ are $\pm\sqrt{n_{f_1}\!+\!n_{b_1}\!+\!
k^2(n_{f_2}\!+\!n_{b_2})}$. In general, the eigenvectors of $H$
are four-fold degenerate (those of $G$ -- resp. two-fold
degenerate), except the ground state (the vacuum $\,|0\rangle :=
|0,0,0,0\rangle)$, which is not degenerate, and the states with
$n_1=0$ or $n_2=0$ (only one mode occupied), which are two-fold
degenerate. We shall use the following basis in the $H$-space:
\beq \ba{lcl}
\psi_1 = |1, n_{b_1}, 1, n_{b_2}\rangle\\[4pt]
\psi_2 = |1, n_{b_1}, 0, n_{b_2}\!+\!1\rangle\\[4pt]
\psi_3 = |0, n_{b_1}\!+\!1, 1, n_{b_2}\rangle\\[4pt]
\psi_4 = |0, n_{b_1}\!+\!1, 0, n_{b_2}\!+\!1\rangle
\ea
\eeq
Any eigenvector $\psi$ of $H$ with eigenvalue $1+n_{b_1}+k^2(1+n_{b_2})$ can
be written as
$$
\psi = \alpha\psi_1 + \beta\psi_2 + \gamma\psi_3 + \delta\psi_4\,.
$$
The eigenvalue-set $(\alpha,\beta,\gamma,\delta)$ determines also
the eigenvectors of $G$:
\beq \ba{ccl} \Phi_1 & = &
\dfrac{1}{\sqrt{2(1+\bar k^2)}}\,(1+\bar
k^2,\,\mp\bar k,\,\pm 1,\,0)\\[8pt]
\Phi_2 & = & \dfrac{1}{\sqrt{2(1+\bar k^2)}}\,(0,\,\pm 1,\,\pm\bar
k,\, 1+\bar k^2)\,, \ea \eeq where $\bar k^2
=k^2(n_{b_2}+1)/(n_{b_1}+1)$. Quantum-mechanical superpositions of
these orthogonal eigenvectors with arbitrary (in general complex)
weights $A\Phi_1 + B\Phi_2$ lead to a density matrix over the
Fermi algebra with eigenvalues \beq \left(\frac{|A|^2}{2},
\frac{|B-\bar k A|^2}{2(1+\bar k^2)}, \frac{|A+\bar k
B|^2}{2(1+\bar k)^2}, \frac{|B|^2}{2}\ \right)\,. \eeq In order to
maximize the entanglement we have to choose $A$ and $B$ such that
$|A+\bar kB|=|B-\bar k A|$, which is achieved for $A=iB$. With
this, the entropy becomes $2\ln2$, so that the corresponding state
is the tracial state over the Fermi algebra. To minimize the
entanglement we have to make $|A+\bar kB|$ and $|B-\bar k A|$ as
different as possible. This is guaranteed for both $A$ and $B$
real, e.g. for $A=\sin \varphi$, $B= \cos \varphi$. In Fig. 1 the
dependence of the entanglement on the mixing parameter $\varphi$
and on the relative weight of the two components $\bar k$ is
shown.

\begin{figure}[ht]
\setlength{\unitlength}{1pt}
\begin{picture}(480,250)
\put(0,0){\makebox(0,0)[lb] 
{\includegraphics*[50,10][540,240]{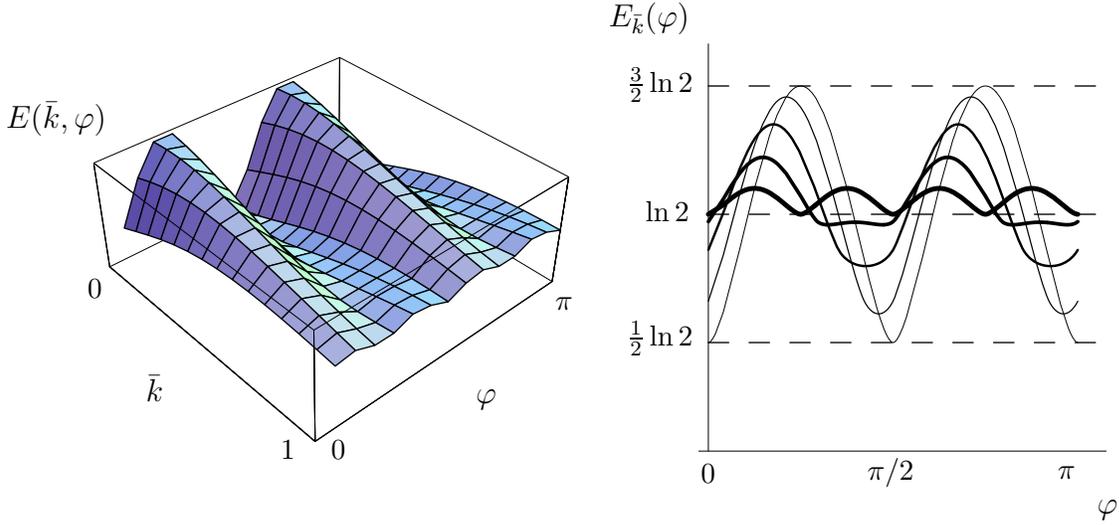}}}
\put(12,170){\makebox(0,0)[l]{$E(\bar k,\vp)$}}
\put(43,106){\makebox(0,0)[l]{{\small 0}}}
\put(65,68){\makebox(0,0)[l]{$\bar k$}}
\put(116,45){\makebox(0,0)[l]{{\small 1}}}
\put(190,64){\makebox(0,0)[l]{$\vp$}}
\put(135,45){\makebox(0,0)[l]{{\small 0}}}
\put(220,100){\makebox(0,0)[l]{{\small $\pi$}}}
\put(247,87){\makebox(0,0)[l]{{\small $\frac{1}{2}\ln 2$}}}
\put(254,135){\makebox(0,0)[l]{{\small $\ln 2$}}}
\put(247,183){\makebox(0,0)[l]{{\small $\frac{3}{2}\ln 2$}}}
\put(240,208){\makebox(0,0)[l]{$E_{\bar k}(\vp)$}}
\put(275,36){\makebox(0,0)[l]{{\small 0}}}
\put(338,36){\makebox(0,0)[l]{{\small $\pi/2$}}}
\put(410,36){\makebox(0,0)[l]{{\small $\pi$}}}
\put(425,22){\makebox(0,0)[l]{$\vp$}}
\end{picture}
\begin{quote}
\caption{Entanglement of the ``susino"-state: {\sl(a)\/} in the complete
parameter range; {\sl(b)\/} for weight factors $\bar k = 0, 0.25, 0.5,
0.75, 1$  (the line-thickness increases with $\bar k$).}
\end{quote}
\end{figure}

The extremal points are then obtained by solving the equation \beq
\ba{rcc} \dfrac{(\cos\varphi - \bar k\sin\varphi)(\sin\varphi +
\bar k\cos\varphi)}{2(1+\bar k^2)}\,\ln\frac{(\cos\varphi - \bar
k\sin\varphi)^2}{(\sin\varphi +
\bar k\cos\varphi)^2} & - & \\[12pt]
\sin\varphi\cos\varphi\,\ln\dfrac{\sin^2\varphi}{\cos^2\varphi} &
= & 0\,. \ea \eeq

For $\bar k =0$ the minimum is achieved for $\sin\varphi\!=\!0$ or
$\cos\varphi\!=\!0$ and amounts to $S(\rho)=\ln 2$. The minimal
entropy increases with $\bar k$ to reach for $\bar k =1$ its
maximal value $3/2\,\ln 2$, as is shown in Fig. 1b.

\begin{figure}[ht]
\setlength{\unitlength}{1pt}
\begin{picture}(480,200)
\put(0,10){\makebox(0,0)[lb] 
{\includegraphics*[60,0][5400,200]{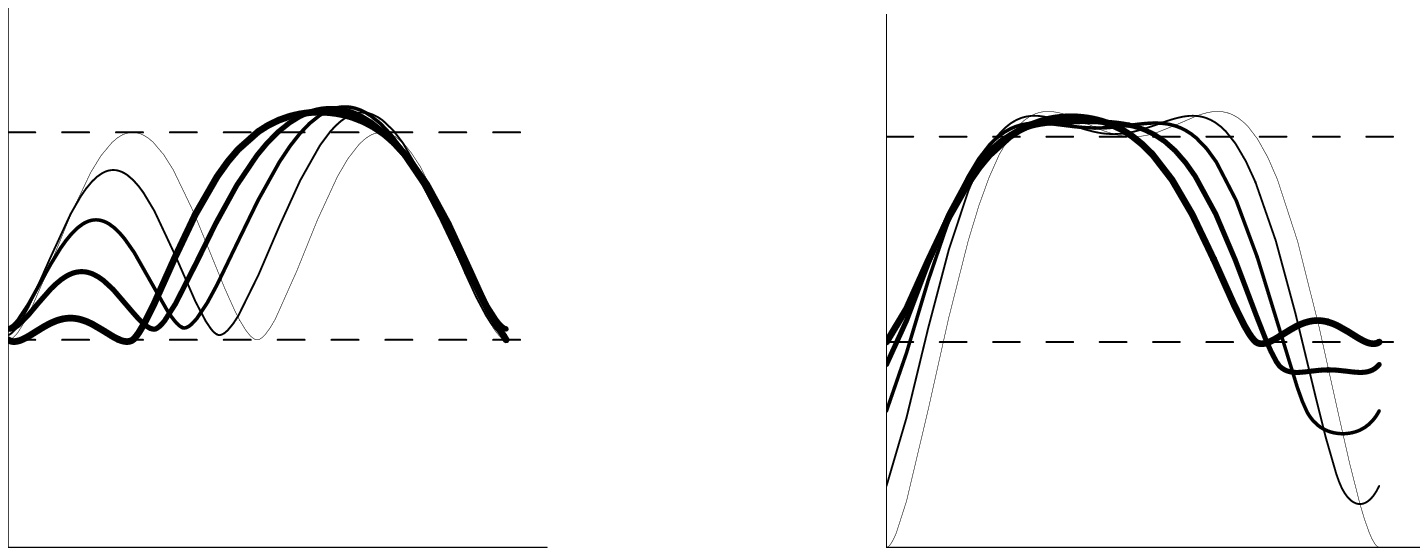}}}
\put(44,10){\makebox(0,0)[l]{{\small 0}}}
\put(110,10){\makebox(0,0)[l]{{\small $\pi/2$}}}
\put(184,10){\makebox(0,0)[l]{{\small $\pi$}}}
\put(200,5){\makebox(0,0)[l]{$\vp$}}
\put(22,80){\makebox(0,0)[l]{{\small $\ln 2$}}}
\put(14,140){\makebox(0,0)[l]{{\small $\frac{3}{2}\ln 2$}}}
\put(2,180){\makebox(0,0)[l]{$E^{(1)}_{\bar k}(\vp)$}}
\put(267,80){\makebox(0,0)[l]{{\small $\frac{1}{2}\ln 2$}}}
\put(275,140){\makebox(0,0)[l]{{\small $\ln 2$}}}
\put(248,180){\makebox(0,0)[l]{ $E^{(2)}_{\bar k}(\vp)$}}
\put(295,10){\makebox(0,0)[l]{{\small 0}}}
\put(355,10){\makebox(0,0)[l]{{\small $\pi/2$}}}
\put(435,10){\makebox(0,0)[l]{{\small $\pi$}}}
\put(450,5){\makebox(0,0)[l]{$\vp$}}
\end{picture}
\begin{quote}
\caption{Entanglement of the first, resp. the second
fermion for weight factors $\bar k = 0, 0.25, 0.5, 0.75, 1$ (the
line-thickness increases with $\bar k$).}
\end{quote}
\end{figure}

Fig. 2 shows the entropy (so, entanglement) of the first,
resp. the second fermion of the susino states for a two-mode
system. Zero entanglement occurs only for $\bar k=0$ when the
second mode is not affected by the supertransformation. However,
$E~=~0$ appears only at two points, otherwise the mere existence  of
the other mode already influences the behavior of the system by
creating some entanglement.

Thus for the entanglement of the eigenstates we find the
following:
\begin{itemize}
\item[{\sl(i)}] The vacuum is not entangled, $E=0$;
\item[{\sl(ii)}] The ``one-mode'' states $n_{b_1}=0$ and $n_{b_2}=0$ are
  characterized with $E=\ln 2$, however the entanglement of the first
  fermion w.r.t. the rest of the system is 0 while for the second
  fermion it is $\ln 2$ and vice versa;
\item[{\sl(iii)}] In the general case the entanglement varies
between its maximal value $E_{max} = 2\ln 2$, which is independent
on $\bar k$, and some minimal value $E_{min}$ already depending on
$\bar k$, for which $\ln 2 \leq E_{min} \leq 3/2\,\ln 2 $.
\end{itemize}

\medskip

\section{KMS-states}

A theorem due to Buchholz and Ojima \cite{BO} says that supersymmetry and
KMS-structure are incompatible. More precisely, they show that an
equilibrium state cannot be invariant under the evolution given by the odd
derivations (1.8). Indeed, (with $x=e^\beta$, $\beta$ being the inverse
temperature)
$$
\ba{lcl}
\omega(a^\dg a) = \dfrac{1}{1+x} & \qquad & \omega(aa^\dg) =
\dfrac{x}{1+x}\\[8pt]
\omega(b^\dg b) = \dfrac{1}{x-1} & & \omega(bb^\dg) = \dfrac{x}{x-1}
\ea
$$
is so different between bosons and fermions that it is hard to believe
that this will not change by mixing them. Nevertheless we shall show that
this happens miraculously if the evolution is governed by $e^{iGs}$. With
the shorthand $(c,s)=(\cos s\sqrt H, \sin s\sqrt H)$, this evolution reads
\beq
a^\dg(s)a(s) = a^\dg a +i\frac{cs}{\sqrt H}[G, a^\dg a] +
\frac{s^2G}{H}[a^\dg a, G]\,.
\eeq
We need consider only $\omega(a^\dg a)$ since $a^\dg a + b^\dg b$ does not
change and if one is invariant so is the other. Now
$$
\ba{l}
[a^\dg a, G] = a^\dg b - b^\dg a\\[6pt]
G[a^\dg a, G] = a^\dg bb^\dg a - b^\dg aa^\dg b
\ea
$$
and
$$
\ba{l}
\omega([a^\dg a, G]) = 0\\[4pt]
\omega(G[a^\dg a, G]) = \dfrac{1}{x+1}\,\dfrac{x}{x-1}
- \dfrac{x}{x+1}\,\dfrac{1}{x-1} = 0\,.
\ea
$$
$G$ is invariant under this supertransformation but $G_\A$ is not and
we still have to verify that its thermal expectation remains
zero. Indeed, with $[G,\,G_\A] = 2i(b^\dg b - a^\dg a - 2a^\dg ab^\dg
b)$, this turns out to be true
$$
\omega([G,\,G_\A]) = 2i\,\left(\frac{1}{x-1} - \frac{1}{x+1} -
2\,\frac{1}{x+1}\,\frac{1}{x-1}\right) = 0\,.
$$
So there remains the question of $c$ and $s$. These are functions
given by convergent series of the form $\dsum_{k=o}^\infty
c_k(s^2H)^k$ and $\Tr e^{-\beta H}H^k A = \partial^k/\partial
\beta^k \, \Tr e^{-\beta H} A$. But since the expectation values
of the additional terms vanish for all $\beta$ the factors
$cs/\sqrt H$, $s^2/\sqrt H$ do not change that and we conclude
\beq \omega(a^\dg(s) a(s)) = \omega(a^\dg a) = \frac{1}{1+x}\,.
\eeq On the contrary with evolution (2.3), as we are dealing not
with automorphisms but only with a one-parameter group of maps, we
get for $Q = a^\dg b$
\beq \omega(Q(s)) = \omega(Q(0)) +
s\,\omega(H) = s\,\omega(H) \not= 0
\eeq
in agreement with the
Buchholz--Ojima theorem.

\medskip

\section{Concluding remarks}

To summarize, we have studied four different transformations of three
mixed Bose--Fermi algebras that do not respect the grading.
In all four cases we have one-parameter groups of
transformations which commute with the time evolution, generated by
$H$. In {\bf Ia} they are automorphisms, in {\bf Ib} only linear and
$^*$-preserving maps, in {\bf II} and {\bf III} they do not transform $\A$ into
$\A$. The case {\bf Ia} represents an explicit form of a nonlinear
transformation of creation and destruction operators which preserves
the CCR/ CAR structure.

\smallskip
A state gives a representation in a Hilbert space and an associated
probability interpretation, therefore an important question to be
discussed is what happens under these transformations with the states.
For the three algebras we get
\begin{itemize}
\item[{\bf I.}] The usual probabilities;
\item[{\bf II.}] SUSY transforms into states of zero norm, so in all
probabilities nothing happens;
\item[{\bf III.}] With nonzero probability SUSY creates the Clifford object
$\theta, \bar\theta$ which actually is unobservable.
\end{itemize}

We have identified the eigenvectors of the unitary implementer of
the supertransformation as SUSY-quasiparticle states (susinos)
with mixed statistics. Except for the vacuum, the susino-states
are entangled. In the degeneracy space of $G$ the entanglement
varies between the maximal possible value and some minimal value
which is bigger than $\ln 2$.

\medskip
Another natural question is the one about the invariant structures.
Though SUSY mixes fermions with bosons, there should be combinations of
them which remain invariant. They can be readily constructed. In fact, one
can find a time evolution commuting with the SUSY transformation such that
only these objects and not the bosons and the fermions are
time-invariant. The situation is analogous to the
$C$-breaking in the $K^0-{\ol K^0}$ system. Finally one can
ask about SUSY-invariant states and in all cases the Fock
vacuum provides such an example. Less trivial is the thermal
distribution of bosons and fermions which
\begin{itemize}
\item[{\it (i)}] is invariant under {\bf Ia};
\item[{\it (ii)}] is not invariant under {\bf Ib}
(the Buchholz--Ojima theorem \cite{BO});
\item[{\it (iii)}] is trivially invariant under {\bf II};
\item[{\it (iv)}] is not invariant under {\bf III}.
\end{itemize}

\medskip
\section*{Acknowledgements}
We thank H. Grosse, E. Langmann, P. Minkowski, H. Nikolai and J. Wess
for the discussions and D. Buchholz and R. Haag for valuable
remarks. N.I. acknowledges the hospitality at the Erwin Schr\"odinger
International Institute for Mathematical Physics, where part of this
work has been completed.

\end{document}